\documentclass[showpacs,prl,floatfix,twocolumn,amsmath,a4]{revtex4}
\usepackage{graphicx}

\begin{document}

\title{Dielectric properties of charge ordered LuFe$_2$O$_4$ revisited: \\ The apparent influence of contacts}

\author{D.~Niermann$^{1}$}
\author{F. Waschkowski$^{1}$}
\author{J.~de~Groot$^2$}
\author{M.~Angst$^2$}
\author{J.~Hemberger$^{1}$}
\email[Corresponding author:~]{hemberger@ph2.uni-koeln.de}

\affiliation{$^1$II.\ Physikalisches Institut, Universit\"at zu K\"oln, D-50937 K\"oln, Germany}
\affiliation{$^2$Peter Gr\"unberg Institut PGI and J\"ulich Centre for Neutron Science JCNS,
JARA-FIT, Forschungszentrum J\"ulich GmbH, 52425 J\"ulich, Germany}

\begin{abstract}
We show results of broadband dielectric measurements on the charge ordered, proposed to be  multiferroic material LuFe$_2$O$_4$. The temperature and frequency dependence of the complex permittivity as investigated for temperatures above and below the charge-oder transition near $T_{CO}\approx 320$~K and for frequencies up to 1~GHz can be well described by a standard equivalent-circuit model considering Maxwell-Wagner-type contacts and 
hopping induced AC-conductivity. No pronounced contribution of intrinsic dipolar polarization could be found and thus the ferroelectric character of the charge order in LuFe$_2$O$_4$ has to be questioned.
\end{abstract}

\date{\today}


\maketitle

In recent years, multiferroics, i.e. materials combining at least two coexisting ferroic order parameters in a single thermodynamic phase, have attracted remarkable interest in condensed matter physics. Currently most promising with respect to application as well as to fundamental aspects is the class of magnetoelectric multiferroics in which ferroelectricity is coupled to magnetism \cite{Spaldin05,Khomskii09}. Such coupling could enable the control of the electric polarization via a magnetic field and of the magnetic order via an electric field. However, the coexistence of ferroelectricity and (ferro-)magnetism needs a certain level of complexity as it may be generated via the interplay of structural and electronic degrees of freedom in transition metal oxides. Among these underlying mechanisms \cite{Khomskii09} two main scenarios for the onset of ferroelectricity may be highlighted: Systems in which ferroelectricity is driven by partially frustrated spiral \cite{Kimura03,Cheong07} or collinear \cite{Dagotto07,Giovanetti11} magnetism and systems in which ferroelectricity arises from complex charge order \cite{Brink08}, discussed e.g.\ for nearly half doped rare earth manganites \cite{Efremov04,Joos07}, nickelates \cite{Giovanetti09}, magnetite \cite{Picozzi09}, or in particular LuFe$_2$O$_4$ \cite{Ikeda05}.
For this latter class of materials the residual conductivity at the charge order (CO) transition, which in addition may be dependent on magnetic and electric fields \cite{Joos07,Sichelschmidt01}, makes it difficult to probe the theoretically predicted onset of ferroelectricity via macroscopic methods like pyrocurrent, hysteretical $P(E)$-loops, or dielectric permittivity measurements. In such cases contact contributions may add capacitive \cite{Lunkenheimer02,Biskup05} or even magneto-capacitive contributions \cite{Catalan06} which will cover the intrinsic sample properties.

The mixed valence (Fe$^{2+}$/Fe$^{3+}$) system LuFe$_2$O$_4$ was proposed to show a novel type
of ferroelectricity based on frustrated charge order within triangular Fe-O-double
layers at $T_{CO} \approx 330$~K \cite{Ikeda05}, which even is proposed to be coupled to magnetism and magnetic field \cite{Subramanian06, Mulders11}. The corresponding ferroelectric moment was suggested to result from a CO configuration of polar bilayers 
with a Fe$^{2+}$/Fe$^{3+}$-unbalance within both sublayers \cite{Ikeda05,Angst08}.
Below the charge order transition magnetic order sets in at $T_N=240$~K, which is altered in a further magneto-structural, first-order type transition at $T_{LT} \approx 175$~K \cite{Xu08}. However, the large permittivity values, magneto-capacitive effects and temperature dependent polarization measurement reported for this material suffer from being influenced by the relatively high residual conductivity. Thus an unambiguous evidence for ferroelectricity by means of dielectric measurements is difficult to give. Schottky-type depletion layers at the contact interfaces or grain boundaries can lead to Maxwell-Wagner effects \cite{Maxwell-Wagner} and hopping conductivity can give a further frequency dependent contribution to the apparent dielectric constant \cite{Lunkenheimer02}. Such effects have already been demonstrated for poly-crystalline samples of LuFe$_2$O$_4$ for temperatures below 300~K and for frequencies up to MHz \cite{Ren11}.
Here we will report on broadband spectroscopic investigations of the permittivity in high quality LuFe$_2$O$_4$ single crystals below and above the CO-transition for temperatures up to 400~K and frequencies up to 1~GHz in order to separate intrinsic and non-intrinsic contributions to the dielectric properties and to elucidate the potentially polar nature of the CO-state.

The single-crystalline samples of LuFe$_2$O$_4$ were grown employing the floating-zone method 
\cite{Christianson08}. Structural and magnetic measurements confirmed the known behavior: in the high temperature phase the samples are hexagonal and show the known sequence of phase transitions at $T_{CO}=320$~K, $T_{N}=240$~K, and $T_{LT}=175$~K. The samples are from the same batch as used for the latest structural studies published in Refs.\ \onlinecite{Angst08} and \onlinecite{Groot12}.
The dielectric measurements were made in a commercial $^4$He-flow magneto-cryostat ({\sc Quantum-Design PPMS}) employing a home-made coaxial-line inset. The complex, frequency dependent dielectric response was measured using a frequency-response analyzer ({\sc Novocontrol}) for frequencies from 1~Hz to 1~MHz. For higher frequencies up to 1~GHz a micro-strip setup was employed and the complex transmission coefficient ($S_{12}$) was evaluated via a vector network analyzer ({\sc Rhode \& Schwarz}). All measurements were performed with the electric field along the crystallographic $c$-axis (the direction for which a spontaneous ferroelectric moment was postulated \cite{Ikeda05}) with a small stimulus 
of the order $E_0\approx 1$~V$_{rms}$/mm. The contacts were applied to the plate-like single-crystals using silver paint in sandwich geometry with a typical electrode area of $A\approx 1$~mm$^2$ and a thickness of $d\approx0.4$~mm.
The uncertainty in the determination of the exact geometry together with additional (but constant) contributions of stray capacitances results in an uncertainty in the absolute values for the permittivity of up to 20~\%. Additional specific heat 
measurements were carried out in a commercial system ({\sc Quantum-Design PPMS}).


\begin{figure}
\centerline{\includegraphics[width=1.0\columnwidth,angle=0]{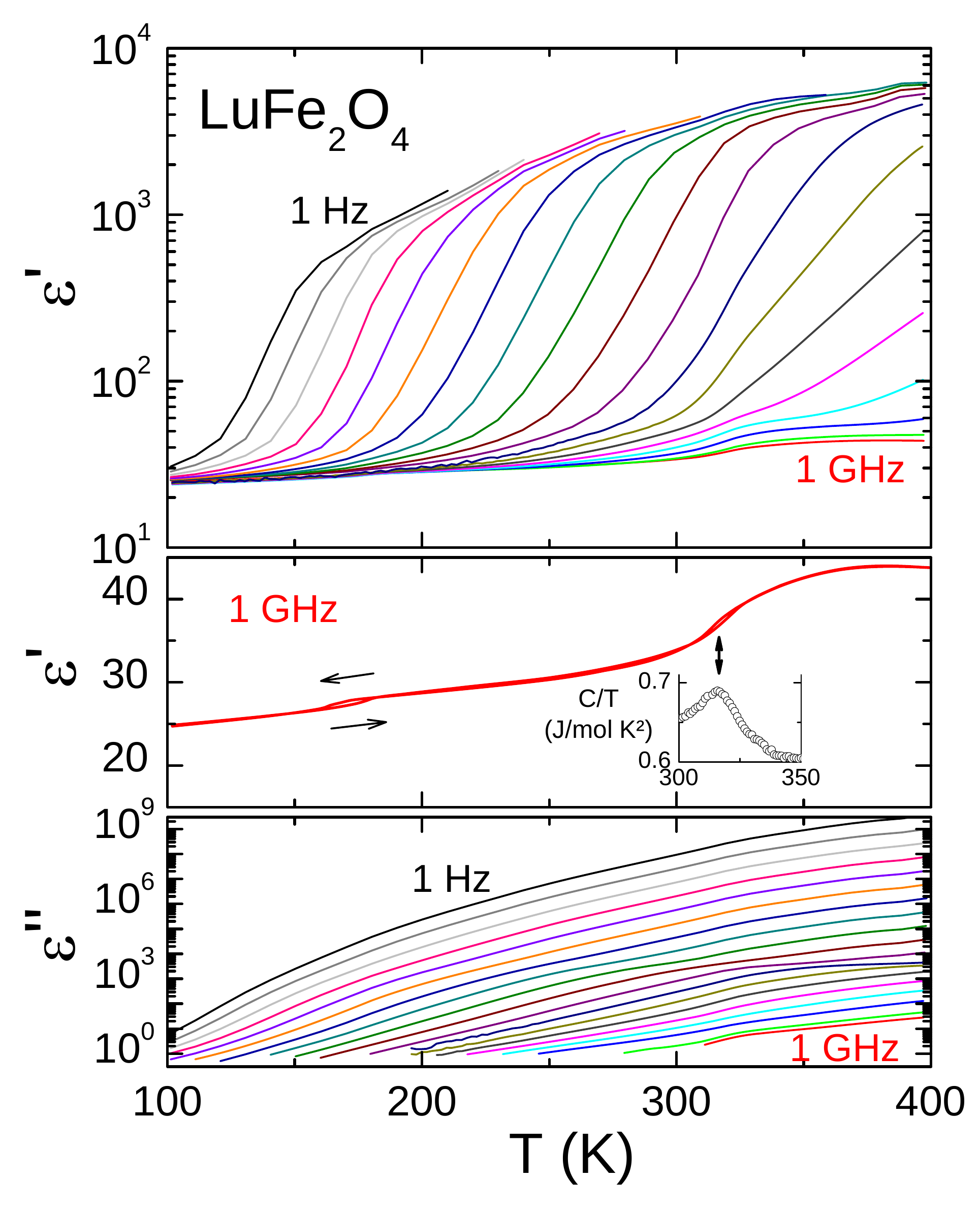}}
\caption{
(Color online) Temperature dependence of the complex dielectric permittivity $\varepsilon^*(T)$ as measured for frequencies between 1~Hz and 1~GHz equally spaced with two frequencies per decade. The middle frame displays the 1~GHz curve of the real part on a linear scale together with specific heat data around the charge order transition.
}
\label{eps(T)}
\end{figure}

Fig.~1 shows the temperature dependence of the complex permittivity for temperatures 100~K~$< T <$~400~K. In the real part $\varepsilon'(T)$ a pronounced step from a low value of roughly $\varepsilon_i \approx 30$ for high frequencies and low temperatures to high values of several thousands for low frequencies and high temperatures. This feature resembles the findings of high permittivity values reported in literature \cite{Ikeda05,Subramanian06}. Already at this point it is remarkable that these high $\varepsilon$-values for low enough frequencies ($\nu < 1 MHz$) do persist for temperatures above $T_{CO}$, and thus obviously do not depend on the onset of possibly ferroelectric charge order. These steps in $\varepsilon'(T)$ are accompanied by cusp-like features in the imaginary part $\varepsilon''(T)$ (Fig.~\ref{eps(T)}, lower frame), which is, however, dominated by a steep, nearly logarithmic, and strongly frequency-dependent increase with temperature. Such type of behavior is due to the influence of conductivity $\sigma$, which in general is via the relation $\sigma'= \omega\varepsilon_0\varepsilon''$ connected to the dielectric loss $\varepsilon''$. The details on these corresponding conductivity contributions will be discussed later but already at this point it shall be mentioned that for high enough frequencies such non-intrinsic features are suppressed (or rather shifted to higher temperatures) and only the intrinsic features persist. Such high frequency data ($\nu=1$~GHz) for $\varepsilon'(T)$ is displayed in the middle frame of Fig.~\ref{eps(T)}, this time on a linear scale. At $T_{LT}\approx 175$~K a small step-like anomaly with a distinct temperature hysteresis is reminiscent of the magneto-structural transition. The magnetic transition at $T_N=240$~K does not show up in the dielectric data, which questions a pronounced magneto-electric coupling. But most remarkably, at $T_{CO}\approx 320$~K no divergent behavior in the permittivity can be detected. In contrast, at the point were the charge order sets in (as reconfirmed via the peak in the specific heat measured on the very same sample, see inset in Fig.~\ref{eps(T)}) $\varepsilon'(T)$ is decreased. This is not compatible with the formation of spontaneous polarization of the order of several $\mu$C/cm$^2$ as reported in literature \cite{Ikeda05}.

\begin{figure}
\centerline{\includegraphics[width=1.0\columnwidth,angle=0]{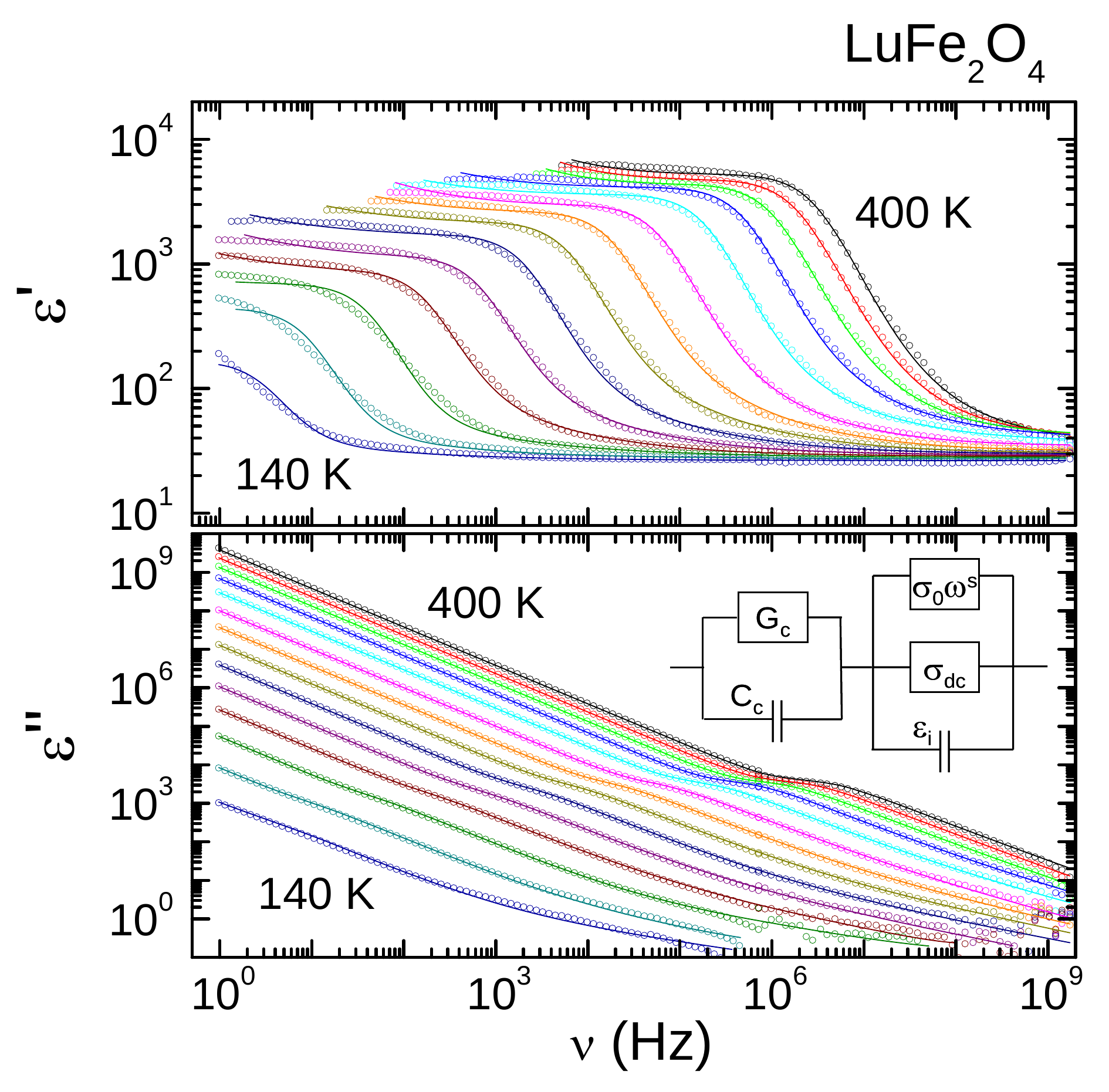}}
\caption{(Color online) Frequency dependence of the complex dielectric permittivity $\varepsilon^*(\nu)$ for temperatures between 140~K and 400~K (in steps of 20~K) and in the frequency range 1~Hz~$\leq \nu \leq$~1~GHz. The data ($\circ$) for the real and imaginary part were fitted simultaneously using the equivalent circuit model described in the text. The fitting results are displayed as solid lines.
}
\label{eps(nu)}
\end{figure}

\begin{figure}
\centerline{\includegraphics[width=0.9\columnwidth,angle=0]{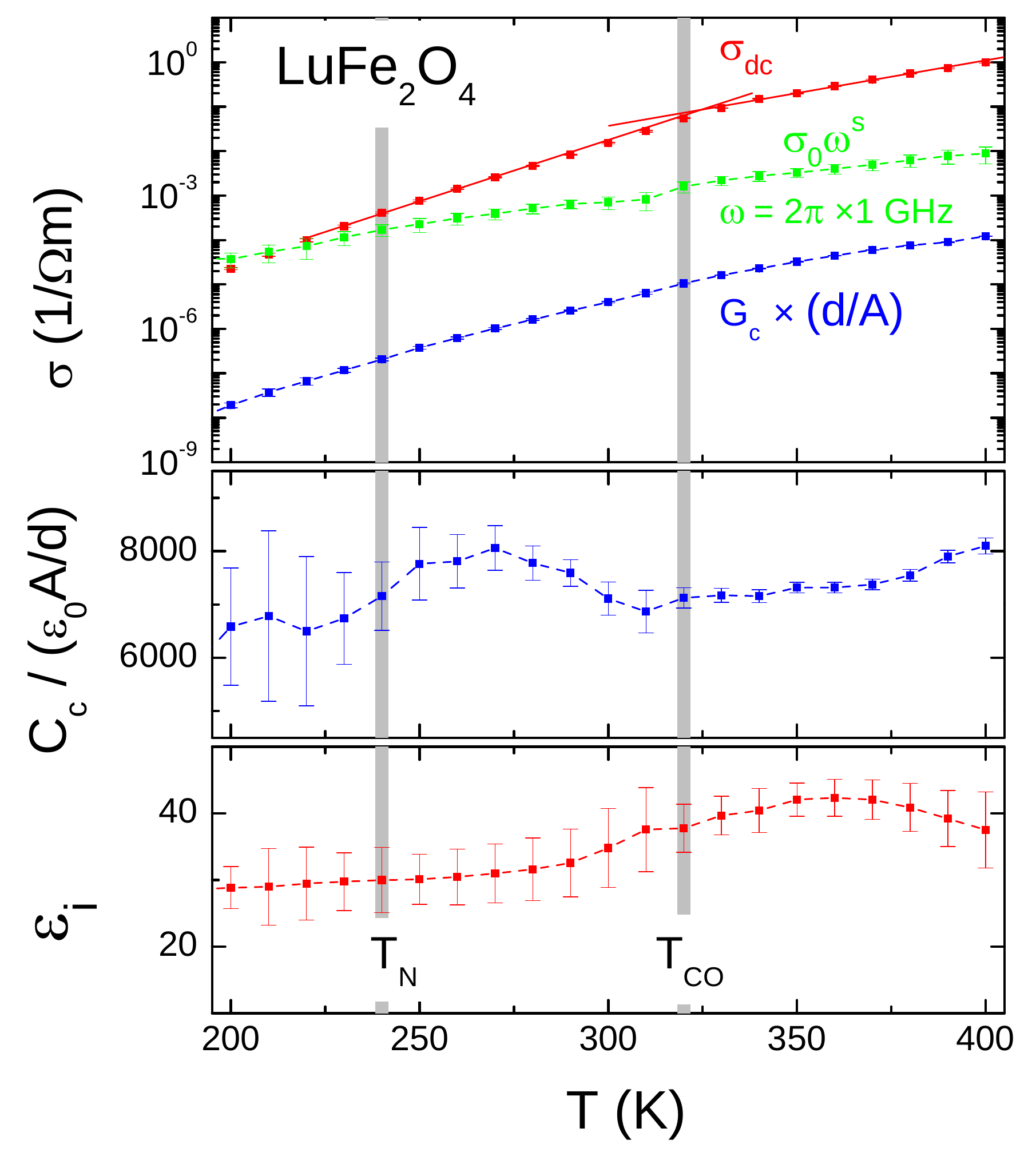}}
\caption{(Color online) Temperature dependence of model parameters gained from the fitting of the permittivity spectra as displayed in Fig.~\ref{eps(nu)}.
The upper frame displays the different conductivity contributions, i.e. the intrinsic DC-conductivity $\sigma_{DC}$, the intrinsic variable range hopping contribution to the conductivity $\sigma_{0}\omega^s$ at a frequency of $\omega = 2\pi\nu=1$~GHz, and the non-intrinsic contribution due to the contact resistance $G_C=1/R_C$ normalized to the sample geometry $A/d$ for reasons of comparability. The solid lines in the $\sigma_{DC}(T)$ curve depict the change of slope near the CO-transition.
The middle frame displays the non-intrinsic capacitive contribution due to the contacts and the lower frame, finally, gives the intrinsic permittivity contribution of the material.
}
\label{parameter}
\end{figure}

In order to shed light onto the origin of the large permittivity values obtained from the dielectric measurements we evaluated the frequency dependent complex permittivity (Fig.~\ref{eps(nu)}). The data roughly can be described as temperature dependent Debye-like steps in $\varepsilon'(\nu)$ accompanied by corresponding peaks in the dielectric loss $\varepsilon''(\nu)$ superimposed by a contribution $\propto 1/\nu$ due to the temperature dependent conductivity. The time constant which defines the step-position for each temperature is given by the effective sample resistance and the contact capacitance $\tau=R(T)C$. The data were quantitatively modeled using an equivalent-circuit description sketched in the lower frame of Fig.~\ref{eps(nu)}. In addition to the intrinsic DC-conductivity $\sigma_{DC}$ and the intrinsic, (for the regarded spectral range) frequency independent permittivity $\varepsilon_i$ of the material the equivalent-circuit model \cite{Lunkenheimer02} contains also the conductance $G_C$ and capacitance $C_C$ of the contacts. In poly-crystalline material additional heterogeneities, e.g.\ grain boundaries, might be considered, which in small single crystals, however, are absent. A further conductivity contribution results from hopping processes in the sample and can be modeled using a frequency dependent term for the ac-conductivity $\sigma_0\omega^s$ (with $\omega = 2\pi\nu$). This term contributes not only to the dielectric loss $\varepsilon''=\sigma'/ (\omega\varepsilon_0)$, but also gives a corresponding Kramers-Kronig consistent contribution to the real part of the permittivity and is commonly described as {\em universal dielectric response} \cite{Jonscher96}. The fits to the data were calculated simultaneously for the real and imaginary part and are displayed as solid lines in Fig.~\ref{eps(nu)}. The data can  convincingly be modeled above and below $T_{CO}$ over the full spectral range of nine decades without the need of further 
contributions 
reminiscent of the onset of ferroelectric order.

The results for the corresponding temperature dependent fitting parameters are displayed in Fig.~\ref{parameter}. The upper frame gives the contributions to the conductivity or the dielectric loss, respectively. The intrinsic DC-conductivity of the sample $\sigma_{DC}$ (red curve in the upper frame of Fig.~\ref{parameter}) shows an approximately exponential decrease with decreasing temperature as expected for semiconductors. Around $T_{CO}$ a change of slope in this semi-logarithmic representation can be assumed (see solid lines in the upper frame of Fig.~\ref{parameter}) reflecting the change of charge carrier mobility at the charge order transition. Similar results were obtained from M\"o\ss bauer-spectroscopy \cite{Xu08}. However, it is remarkable, that the contribution of the contact resistance $1/G_C$ (blue curve in the upper frame of Fig.~\ref{parameter}, displayed as normalized on the sample geometry) dominates the different contributions at all temperatures. The hopping contribution (green curve in the upper frame of Fig.~\ref{parameter}) is displayed as $\sigma_{0}\omega^s$ for $\omega/2\pi=1$~GHz. Again near $T_{CO}$ a small anomaly can be detected. The parameter $s$ possesses a weak and monotonic temperature dependence around values close to $s \approx 0.6$ in agreement with canonical expectations \cite{Lunkenheimer02,Jonscher96}.
The middle frame gives the results for the non-intrinsic contact capacitance $C_C$ displayed as contribution to the ``effective'' dielectric constant, i.e.\ normalized on the geometric capacitance of the sample. The large values of about $\approx 7000$ are more or less constant within the error bars which strongly increase for lower temperatures as the corresponding relaxational step shifts more and more out of the regarded frequency range. Such a weakly temperature dependent capacitance contribution can be understood in terms of very thin depletion layers formed by the Schottky-type metal-semiconductor interfaces at the electrodes. This contribution dominates the capacitive response of the sample in the low frequency, high temperature regime.  The intrinsic contribution to the dielectric constant $\varepsilon_i(T)$ is displayed in the lower frame of Fig.~\ref{parameter}. The residual values lie between 30 and 40, comparable to other transition metal oxides \cite{Lunkenheimer02} but far from the large ``effective'' values generated by the contacts. The curvature of $\varepsilon_i(T)$ corroborates the data obtained for high frequencies as displayed in the middle frame of Fig.~\ref{eps(T)}. Again the decrease of the permittivity for crossing $T_{CO}$ into the CO-phase and the absence of any divergent characteristic at the transition temperature does not point towards the onset of ferroelectricity. This interpretation meets recent results of structural refinements of x-ray diffraction data from the charge-ordered phase of LuFe$_2$O$_4$ where the polar character of the bilayers could not be verified \cite{Groot12}. 
Also scenarios in which disorder smears out the onset of spontaneous polarization and relaxor ferrolectric behavior emerges can be ruled out as explanation for  the relaxational features found in the dielectric response of LuFe$_2$O$_4$. Such a scenario has been proposed e.g.\ for the charge ordered phase of magnetite \cite{Schrettle11}, but then the corresponding relaxation strengths should increase towards lower temperatures while in the present case the effective relaxation strengths decreases in accordance with the interpretation of an origin due to contacts and hopping conductivity. In addition, we repeat that such a strongly conductivity dominated scenario may explain the reported anomalies in pyro-current measurements or the $P(T)$ data derived from them \cite{Maglione08}: Charges are trapped inside the ``hetero-structure'' of contacts and sample for low conductivity values at low temperatures and released when the conductivity is enhanced at higher temperatures close to the CO-transition.

Summarizing, we performed broadband dielectric spectroscopy on single crystalline LuFe$_2$O$_4$ in the frequency range 1~Hz~$<\nu<$~1~GHz for temperatures well above and below the charge order transition at $T_{CO} \approx 320$~K. The results for the frequency and temperature dependent complex permittivity can be modeled quantitatively in terms of extrinsic contact contributions and intrinsic contributions due to finite DC-conductivity, hopping induced AC-conductivity, and intrinsic dielectric permittivity. The results for the intrinsic dielectric properties do not posses any features reminiscent of the onset of ferroelectric order. Thus we suggest to reconsider the polar nature of the charge ordered state in LuFe$_2$O$_4 $. In order to elucidate the ordering phenomena in this interesting but complex 
system further experimental and theoretical investigation are highly desirable.


This work has been funded by the DFG through SFB608 (Cologne).
Support from the initiative and networking fund of
Helmholtz Association by funding the Helmholz University
Young Investigator Group ``Complex Ordering Phenomena
in Multifunctional Oxides'' is gratefully acknowledged. MA thanks D.~Mandrus, B.C.~Sales, W.~Tian and R.~Jin for their assistance during sample-synthesis, also supported by US-DOE.


\end{document}